\documentclass{article}

\pdfoutput=1
\usepackage[utf8]{inputenc} 
\usepackage[T1]{fontenc}    
\usepackage[colorlinks=true,linkcolor=blue, citecolor=blue, urlcolor=blue]{hyperref}       
\usepackage{url}            
\usepackage{booktabs}       
\usepackage{amsfonts}       
\usepackage{nicefrac}       
\usepackage{microtype}      
\usepackage{graphicx}
\usepackage{comment}
\usepackage{lmodern,textcomp}
\usepackage{subcaption}
\usepackage{amsmath}
\usepackage{amsbsy}
\usepackage{longtable}
\usepackage{lscape}
\usepackage{verbatim}
\usepackage{float}
\usepackage{booktabs}
\usepackage{dcolumn}
\usepackage{comment}
\usepackage{multirow}
\usepackage{amsthm,amsmath}
\usepackage{tabularx}
\usepackage{authblk}
\usepackage[section]{placeins}
\usepackage{natbib}
\setcitestyle{authoryear,open={(},close={)}}
\graphicspath{ {Figures_SN/} }

\title{Inclusive Universities.\\
\large Evidence from the Erasmus Program}

\author[1]{ Luca De Benedictis}
\author[2]{ Silvia Leoni}

\affil[1]{\small University of Macerata and LUISS, Italy, luca.debenedictis@unimc.it }
\affil[2]{\small University of Leicester, UK, silvia.leoni@leicester.ac.uk}

\begin{document}

\maketitle

\begin{abstract} 
The Erasmus Program is the main international mobility program in Europe and worldwide. Since its launch in 1987, it has been growing both in terms of participants and budget devoted to its activities. However, despite the possibility to obtain additional funding, the participation of students with special needs to the program remains extremely low. This work quantifies the participation of these students to Erasmus and explores the network of universities involved in their mobility, along the period 2008-2013. In addition, it proposes a novel index to measure the level of inclusiveness of universities welcoming international students with disabilities. Quantifying and analysing this aspect could be the basis for better designing targeted policies and for widening the participation of students with impairments to international mobility.
\end{abstract}





\section{Introduction}

The Erasmus Program (European Region Action Scheme for the Mobility of University Students) is the most famous example of student mobility in Europe and probably worldwide. Since its launch in 1987, the program has grown steadily, contributing to internationalize the higher education (HE) path of millions of students. In 2014, it transformed into Erasmus+ allowing also young people not in education, teaching and administrative staff to take a period of mobility abroad for training, teaching or carrying out activities within EU relevant projects. Its implementation is not limited to program countries in Europe, but it is extended to partner countries across the world, making it unique for reach and reputation. Its success is confirmed by the recent launch of the new Erasmus+ Program 2021-2027 with a budget of €26.2 billion, compared with €14.7 billion for 2014-2020 \citep{EuroCommission}.\footnote{\footnotesize The success of Erasmus has been recently celebrated by awarding a Honorary doctorate in Social, Developmental and Educational Psychology to Sofia Corradi, the creator of the Erasmus Program, by the Sapienza University in Rome \citep{Sapienza}.} The increased budget should allow for a more inclusive, more digital, and greener Erasmus. 

However, its importance worldwide is not associated to an equal relevance in the scientific literature. Beside the internal reports produced by the EU institutions \citep{european2015erasmus+}, literature has overlooked the participation to the Erasmus mobility and, as a consequence, little is known about its structure and evolution over time. 
Even less is known about the participation of students with special needs, which would deserve major focus. Missed participation by students with disabilities due to barriers to the mobility may lead to potential loss of individual and social benefits. 

Erasmus' legal basis emphasize the need to widen the program to people belonging to under-represented groups, or with special needs or fewer opportunities \citep{eu-1288-2013}.
In particular, the program commits to ensure the participation of individuals with physical, mental or health-related conditions by providing specific attention to guidance and accessibility, as well as additional funding via the Erasmus+ special needs support. When preparing for the Erasmus mobility, participants who wish to request these funding, need to indicate their extra costs following the application procedure established by each higher education institution. The additional grant is provided to offset specific difficulties faced by the participant, such as adapted accommodation, accompanying person, supportive equipment, adaption of learning material.

Nonetheless, Erasmus reports indicate an extremely limited participation by students with special needs \citep{european2015erasmus+} and, contextually, the literature has not explored and quantified this phenomenon. This gap may be related to the scarcity and inaccuracy of statistical data available for international students with disabilities, as claimed by \cite{du2018designing}.

This work intends to contribute to filling  this gap by exploring i) the participation of students with special needs in Erasmus for study abroad and ii) the level of inclusiveness of participating universities. The ultimate objective is to increase awareness of potential issues related to this phenomenon and provide a quantitative basis for bodies in charge of related policies. 
We contribute to the literature in three ways.

First, we quantify the participation of students with disabilities in the Erasmus program for study reason only, with respect to students with special needs enrolled in HE in Europe, by exploring mobility by country and gender. Findings show that an extremely low share of Erasmus students is represented by students with disabilities and even a smaller portion of students with special needs in HE participate to the mobility. Almost every country participating to the program in 2013 showed a higher share of female students with disabilities leaving for the mobility.

Second, adopting the tools of Social Network Analysis, we analyze the network of universities participating in the Erasmus mobility of students with special needs and compare it with the network of universities related to the overall participation of Erasmus students, explored in prior research \citep{de2020gender}. We explore the topology of the network and verify whether the bias in favour of women which characterizes the overall network \citep{maiworm2001erasmus,bottcher2016gender,de2020gender} is also present in the subnetwork of participants with special needs. This subnetwork appears much sparser than the overall network. The gender bias persists and increases along the period 2008-2013, contrary to the tendency found in the overall network. In the years of reference, an increasingly low share of participants study a STEM discipline. Universities involved in the mobility of students with disabilities polarize in the role of senders or receivers, with the exception of universities located in countries' capital cities, which act as both senders and receivers. In particular, sending universities are located in Italy, Germany and Eastern countries, whereas receiving institutions follow a South-West North-East axis, including Spain, France, UK, and Northern countries such as the Netherlands and Sweden. 

Third, we propose a novel index to measure the level of inclusiveness of HE institutions participating in Erasmus. Only 13 universities hosted Erasmus students with special needs in every year between 2008 and 2013 and among those some institutions outperform the average level of inclusiveness of their respective country, e.g. the University of Oslo and the Rijksuniversiteit Groningen, while others show a lower level of inclusiveness with respect to their country average, e.g. the University of Valencia and the Polytechnic University of Valencia. 

The paper is organized as follows: the following section describes current research on the topic of international mobility of students with disabilities; the next section provides a quantification of the Erasmus mobility of students with special needs with respect to the number of students with disabilities in HE, and considers differences by gender; we then provide results from the analysis of the network of universities involved in the mobility of students with disabilities and finally we explore the level of inclusiveness of a group of universities, selected as the most inclusive for the period considered. We conclude with a summary of the main findings and a discussion on the limitations of our study and its potential extension in future research.

\section{Prior research}
\subsection{Benefits and barriers to international mobility}
Research showed that international mobility contributes to students' personal development \citep{keogh2009exchange} through improved problem solving skill \citep{behrnd2012intercultural}, better knowledge of foreign languages \citep{otero2006survey}, more self-confidence \citep{braskamp2009assessing}, increased autonomy and flexibility \citep{kitsantas2004studying,papatsiba2005political}, and future employability \citep{bryla2015impact,engel2010impact,parey2010studying,d2021international}, as well as cultural awareness and the formation of individual identity  \citep{oborune2013becoming,langley2005interacting,teichler2001mobility}. For the case of Italy, \cite{d2021international} found that international mobility is linked with a higher probability to enroll in postgraduate studies.
Improvement in soft skills and future career opportunities is proved also for students with disabilities \citep{hameister1999college}, for which better employability becomes particularly significant given the higher unemployment rate in their group \citep{disabilitystat}. In addition, international mobility has showed to have specific advantages for disabled students. \cite{shames2005impact} found that after studying abroad, students with learning disabilities (LDs) and attention deficit hyperactivity disorder (ADHD) reported increased intellectual curiosity and more active engagement in the academic coursework and with peers, improved self-confidence and improved knowledge of physical space and ability to orientate themselves, which can be difficult due to their disability.

Besides the potential benefits, prior research that has dealt with the international mobility of students with disabilities explored the barriers to participation and the best practices and policies that single HE institutions and countries shall adopt. 
\cite{heirweg2020study} conducted a study on 74 students with disability at the University of Bologna in Italy and found that they encountered financial, linguistic and technical barriers (related to study programs and recognition of credits) in line with findings for barriers encountered by their non-disabled peers \citep{souto2013barriers, doyle2010investigation}. In addition, they met practical barriers concerning finding accommodations abroad and building
a social life, and they claimed a lack of sufficient information about the support available at the host university, in line with findings by \cite{du2018designing} for the case of South African HE institutions. \cite{johnstone2020accommodations} argue that efforts by HE institutions in favor of an increased level of accessibility are focused on providing appropriate accommodations to international students with disabilities but they are still at the beginning of including an accessibility culture into the design of study programs.

\subsection{The Erasmus mobility as a network}
Whereas current research has adopted qualitative methods, our approach is quantitative and relies on social network analysis to highlight the structure of students' flows.
The network approach is not new to the study of international student mobility \citep{shields2013globalization} and Erasmus in particular \citep{restaino2020analysing,breznik2020erasmus,breznik2017institutional,breznik2016erasmus,derzsi2011topology,breznik2015exploring}. Prior research mostly conducted analysis at country level rather than university level, with the exception of \cite{de2020gender} who focus on differences by gender. Research has provided an overview of the most active sending and receiving countries \citep{breznik2020erasmus,restaino2020analysing} as well as of the topology of the Erasmus network \citep{derzsi2011topology}, by considering the network of all participants, without specific concern for students with disabilities. 

On the one hand, social network analysis has not been employed to study the international mobility of disabled students, but on the other hand it is difficult to find examples of its application to study the participation of students with disabilities in HE in general.
Social network analysis has been adopted though to study friendship networks and peer acceptance of students with disabilities in primary and middle school, with results with limited external validity and generalizability \citep{mamas2020employing,mamas2020friendship,de2013peer,farmer1999social}. 

\section{Erasmus for all?}
To quantify the participation of students with disabilities to Erasmus we rely on freely accessible data available at the EU open data portal. Notwithstanding Erasmus' long history, they consist of a limited number of datasets corresponding to different academic years and containing information for each participant to the mobility. We limit our analysis to the six-year period between 2008 and 2013, for a homogeneous comparison with the network examined in \cite{de2020gender} and we rely on data for 2008-2018 only for aggregate figures.
The information provided includes the type of mobility (study or placement), the home country and the host country, the home university and the host university, the field of study, the participant's gender, and whether the participant received additional funding for special needs, coded as a binary variable or reporting the amount of extra funding, depending on the dataset. We limit our analysis to participants involved in student mobility and discard observations related to traineeships and staff mobility. We select participants who benefited from the additional special need financial support, thus observations which report a ``yes'' or a value different than zero for the variable ``special needs''. This information may not exactly correspond to the number of students with disabilities involved in the mobility as it depends on self-disclosure of one's health condition and on the request for the extra grant \citep{bound2001measurement}, but we assume that it can be considered a good proxy for the number of disabled students involved, given the personal and national socio-demographic factors that can influence the individual propensity to apply to special needs funding.

When announcing the next Erasmus program in 2011, the full name of the program was ``Erasmus for all'' to recall its inclusive nature.
Yet, despite its known benefits and the extra grant provided by the EU, the participation of disabled students in the program remains low, signaling that additional funding do not compensate for the barriers that these students may encounter when undertaking the mobility.

\begin{figure}[h]
      \includegraphics[width=12cm, height=8 cm]{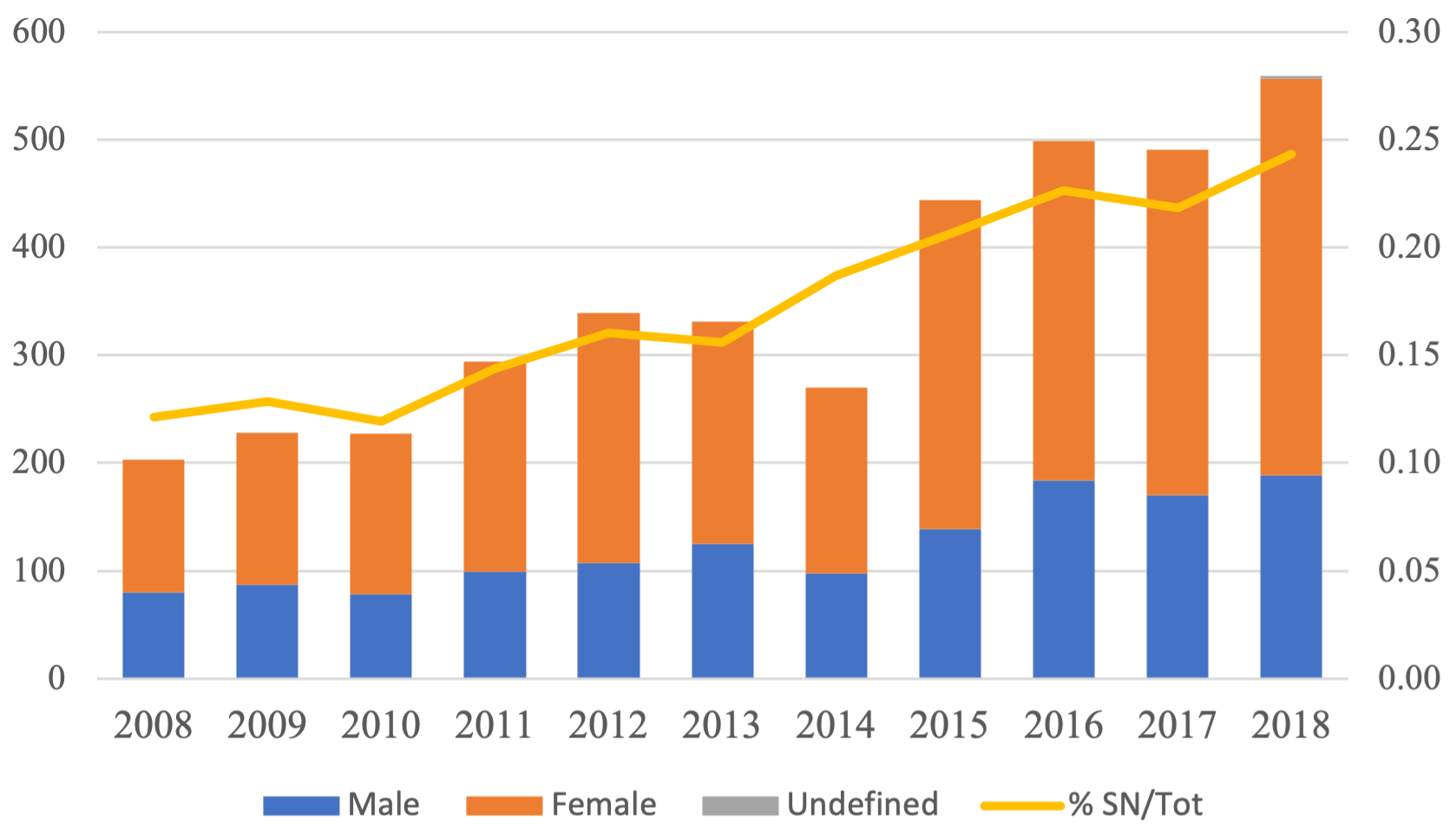}
       \caption{Evolution of the participation of students with special needs to Erasmus.\\
       The bars indicate the number of participants for each year (to be read on the left vertical axis). The yellow line represents the share (\%) of students with special needs over the total number of Erasmus students (to be read on the right vertical axis).
      }
      \label{fig:fig1}
\end{figure}

In 2018 only the 0.24\% of Erasmus students received a financial support for special needs. This low figure may reflect the low participation rates of students with special needs in HE and their high rate of drop out from education \citep{disabilitystatedu}. However, the number of participants with disabilities has more than doubled between 2008 and 2018 following the general trend in the overall participation and their rate of participation over the total number of participants has doubled with respect to its value in 2008 (0.12\%). This growth is displayed in Fig. \ref{fig:fig1} for the period 2008-2018 with bars for the absolute number of participants and with a yellow line indicating the percentage ratio between students with special needs and total students including disabled and non-disabled (\% SN/Tot); the figure also shows participation by gender and confirms that the known gender bias in favor of female Erasmus participants \citep{de2020gender} persists when considering only students with special needs. However, while the overall flows show a mild tendency towards reduction of this bias in latter years, the flows of students with special needs show an increase of the ratio between female and male participants which is equal to 1.54 in 2008 and 1.95 in 2018.
\medskip

\begin{figure}[h]
      \includegraphics[width=12cm, height=8 cm]{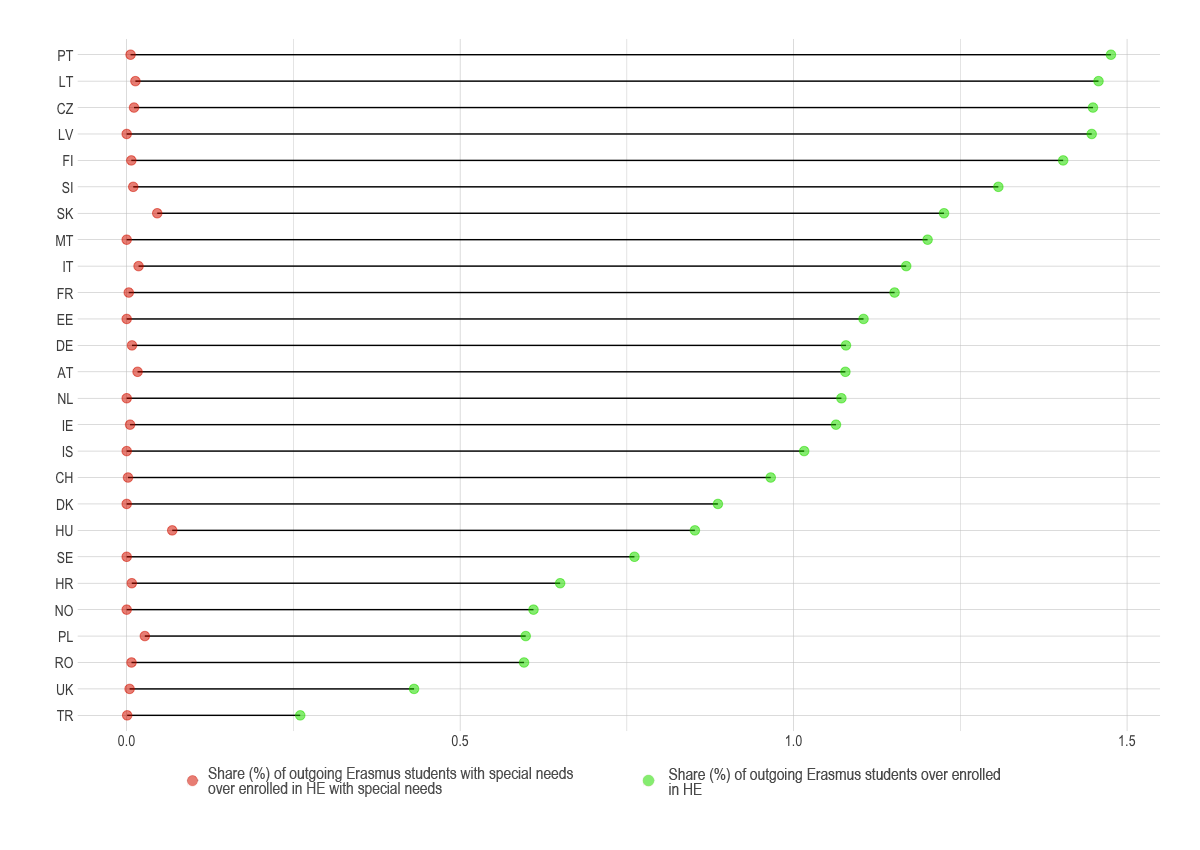}
      \caption{Share of Erasmus students over total population of students in HE in 2013.\\
       Labels correspond to ISO alpha-2 country codes. Countries are displayed in descending order by share of Erasmus students over enrolled in HE (green dots). Belgium, Bulgaria, Cyprus,
       Spain, Greece, Liechtenstein, Luxembourg, Republic of Macedonia are not displayed due to missing information on the number of students with impairments in HE. Information on the population of students with disabilities in HE for the UK was retrieved from the Higher Education Statistics Agency for the year 2014.} 
      \label{fig:fig2}
\end{figure}

To understand the order of magnitude of the participation of students with special needs to Erasmus with respect to those enrolled in HE, we attempted to collect data on students with disabilities in HE in the European countries by contacting the ministry of higher education of each country participating to the program. However, we received inadequate response, with only a few countries keeping a systematic record of enrolled students with disabilities, hardly differentiating among type of disability. Moreover, when present, the data collection process appears to have been established only in the latest years and the dishomogeneity of data across country suggests that a country comparison would not be reliable. The lack of accurate information about students with disabilities in HE could depend on the protection of sensitive data, unwillingness to disclose personal details about health condition or insufficient awareness and communication among the players involved \citep{de2013peer}.
Therefore, we turned to information provided by Eurostudent, a project carried out by a consortium of organizations led by the German Centre for Higher Education Research and Science Studies (DZHW), with the aim to provide cross-country comparison of data on the social dimension of European higher education to support researchers and policy-makers. Information comes from the sixth round of the project to which 28 countries of the European Higher Education Area (EHEA) have contributed between 2016 and 2018, and it provides the share of students with impairments in HE on the basis of surveys submitted in the countries of interest.\footnote{\footnotesize Students with impairments include all students with long-standing health problems, and functional limitations (physical chronic disease, mental health problems, mobility impairments, sensory impairments (vision and hearing), and learning disabilities or other) regardless of the impact on their studies/ everyday life activities.} 
We apply the shares identified by the Eurostudent surveys on the number of students in HE provided by Eurostat for the year 2013 and in Fig. \ref{fig:fig2} we compare the share of outgoing Erasmus students by country over the total student population in HE for the case of students with disabilities only and for disabled and non-disabled students together. The share of students in HE leaving for the mobility is quite low, it never reaches 1.5\%, with heterogeneous values across countries. On the other hand, the share of students with disabilities undertaking the mobility is extremely lower, with values close to zero. To provide an example, in Germany, only 51 students left for the mobility out of approximately 643000 estimated students with disabilities in HE; in France, 7 students left for the mobility out of approximately 224000 estimated students with disabilities in HE. Hungary and Slovakia report the highest shares but all countries show very small percentages. These values are displayed in Table \ref{tab:tab1} with a breakdown by gender. 
The share of women with disabilities participating in the mobility is larger in almost every country, with Hungary showing the highest value. In 2013 Hungary sent abroad 26 female students and 12 male students with disabilities.
Exceptions are represented by Croatia, France, Italy, and Switzerland which sent abroad a higher share of male students with special needs. Countries such as Denmark, Estonia, Iceland, Latvia, Malta, Norway, Sweden, The Netherlands had no outgoing Erasmus students with disabilities in 2013.

The limited availability of systematic and detailed data on the topic only allows to speculate on the reasons why there is a prevalence of female outgoing Erasmus student. \cite{de2013peer} showed that in school it is less likely for girls with disabilities to be accepted by their female peer when they show social problems. We hypothesize that if this characteristic persists in HE, it could represent a push factor for the mobility in view of a potential greater acceptance abroad. In addition, there could be a gender difference in the ability to face barriers to mobility, and/or a need for a superior investment in international education to compensate the foreseen gender gap in the labor market.

\begin{table}[htbp]
  \centering
  \caption{Share of outgoing Erasmus students with disabilities over male, female and overall students with disabilities in HE in 2013, by country and by gender.}
    \begin{tabular}{lrrr}
    \hline
    Country & \multicolumn{1}{l}{M \%} & \multicolumn{1}{l}{ F \%} & \multicolumn{1}{l}{overall \%} \\
    \hline
    Austria & 0.0037 & 0.0264 & 0.0164 \\
    Croatia & 0.0094 & 0.0064 & 0.0077 \\
    Czech Republic & 0.0072 & 0.0140 & 0.0110 \\
    Denmark & 0.0000 & 0.0000 & 0.0000 \\
    Estonia & 0.0000 & 0.0000 & 0.0000 \\
    Finland & 0.0000 & 0.0116 & 0.0069 \\
    France & 0.0051 & 0.0016 & 0.0031 \\
    Germany & 0.0073 & 0.0086 & 0.0079 \\
    Hungary & 0.0487 & 0.0841 & 0.0684 \\
    Iceland & 0.0000 & 0.0000 & 0.0000 \\
    Ireland & 0.0000 & 0.0096 & 0.0051 \\
    Italy & 0.0245 & 0.0135 & 0.0178 \\
    Latvia & 0.0000 & 0.0000 & 0.0000 \\
    Lithuania & 0.0129 & 0.0133 & 0.0131 \\
    Malta & 0.0000 & 0.0000 & 0.0000 \\
    Norway & 0.0000 & 0.0000 & 0.0000 \\
    Poland & 0.0236 & 0.0292 & 0.0271 \\
    Portugal & 0.0052 & 0.0063 & 0.0058 \\
    Romania & 0.0048 & 0.0095 & 0.0072 \\
    Slovakia & 0.0404 & 0.0488 & 0.0457 \\
    Slovenia & 0.0000 & 0.0157 & 0.0099 \\
    Sweden & 0.0000 & 0.0000 & 0.0000 \\
    Switzerland & 0.0047 & 0.0000 & 0.0020 \\
    The Netherlands & 0.0000 & 0.0000 & 0.0000 \\
    Turkey & 0.0000 & 0.0016 & 0.0008 \\
    United Kingdom & 0.0043 & 0.0045 & 0.0044 \\
    \hline
    \end{tabular}%
  \label{tab:tab1}%
  
  {\scriptsize {\bf Note}: {\raggedright \linespread{0.5}\selectfont 
  M stands for male; F stands for female. 
  Belgium, Bulgaria, Cyprus, Spain, Greece, Liechtenstein, Luxembourg, Republic of Macedonia are not displayed due to missing information on the number of students with impairments in HE. Information on the population of students with disabilities in HE for the UK was retrieved from the Higher Education Statistics Agency for the year 2014.\par}} 
  
\end{table}%

\section{A network of inclusive universities}

Following \cite{de2020gender}, we explore and visualize the network of universities taking part in the exchange program of students with disability. This represents a subnetwork of the network analyzed in \cite{de2020gender}. To allow for a comparison we maintain the focus on years 2008 and 2013 and separate the analysis by gender. 

The Erasmus Network at time $t$, $\mathcal{N}_{t}$ is a one-mode network defined by four sets of elements: $\mathcal{N}_{t}=(\mathcal{V}, \mathcal{L},\mathcal{W},\mathcal{O})$, where $\mathcal{V}$, identifies a set of nodes, represented by HE institutions; $\mathcal{L}$ is a set of directed arcs which identify the existence of an Erasmus exchange program between universities;  $\mathcal{W}$ is the edge value function containing the weights corresponding to the flows of students involved in the mobility; $\mathcal{O}$ is the node value function containing information on universities, and on the country they belong.
We first analyze the unweighted directed Erasmus network corresponding to a binary adjacency matrix $\mathcal{A}_{t}$, containing elements $a_{ij}=0$ if university $i \in \mathcal{V}$  does not link to university $j\in \mathcal{V} $ and $a_{ij}=1$ otherwise. Summary statistics are reported in Table \ref{tab:tab2}. The network appears to be much more restricted with respect to the network including also non-disabled students. It is made of 901 academic institutions with 252 active universities in 2008 (with 649 \texttt{Isolates}), out of the 2290 universities participating to the Erasmus program, and 388 in 2013 (and 513 \texttt{Isolates}, giving evidence to a low degree of persistence in the participation of universities in the network)\footnote{\footnotesize Many universities are active senders or receivers only from time to time, given the very low number of students with disability that take advantage of the Erasmus program.}, out of 2658 universities. The data on active universities shows relevant characteristic of the Erasmus network of students with disabilities: the prevalence for \texttt{receiving} rather than \texttt{sending} students in mobility. This brings up the existence of a cluster of \emph{inclusive universities}, hosting students with disabilities in higher proportions. Comparing the two years under scrutiny, $\mathcal{L}_{2008}=199$ and $\mathcal{L}_{2013}=324$ indicates that 125 more partnerships were established between 2008 and 2013. Like the overall network, it is characterized by numerous isolated nodes whose number decreases along time, as its \texttt{Density} increases. However, the network remains quite sparse, much more than the corresponding overall network in \cite{de2020gender} (see the values of the \texttt{Density} in squared brackets), signaling that the probability to have a tie between two random nodes is just a little bit higher than zero. \texttt{Degree} centralization and \texttt{Closeness} centralization present values close to zero signaling that the network is far from having a hierarchical structure. The level of homophily in the network increases over time with a modest disassortativity in 2008 changing into a modest level of \texttt{Assortativity} in 2013, meaning that universities increasingly tend to connect with other universities showing similar characteristics in connectivity. However, the level of reciprocity, i.e. the likelihood of nodes to be mutually connected, is null. All this shows that the network is somehow randomly formed without any systematic attempt to coordinate an international university policy to favor the international mobility of HE students with disabilities.

Looking at differences by gender, in line with findings from \cite{bottcher2016gender} and \cite{de2020gender}, the female Erasmus network of students with disabilities is more connected than the male network, with a ratio between $\mathcal{L}_{2008}^{F}$ and $\mathcal{L}_{2008}^{M}$ equal to 1.525 and between $\mathcal{L}_{2013}^{F}$ and $\mathcal{L}_{2013}^{M}$ equal to 1.624. The bias in favor of women persists and increases over time, contrary to the mild tendency to a reduction shown by the overall network. The male network is characterized by a modest disassortativity in both years considered, whereas the female network aligns to the tendency towards an increased assortativity along time, showed by the whole network of Erasmus students.

We also explored the weighted version of the network, taking into account flows of students associated to each link and computed the strength of the network as the sum of all weights, by gender and field of study aggregated in STEM and non-STEM disciplines.\footnote{ \footnotesize Fields are classified according to the ISCED-F 2013 classification. The STEM fields include \texttt{Engineering, manufacturing and construction}, \texttt{ICTs} and \texttt{Natural sciences, mathematics and statistics}. The remaining fields are classified as non-STEM.}. Fist of all, the small difference between \texttt{University partnerships} and \texttt{Active connections} indicates that sending universities tend to send one single student with special needs (the rare exceptions are quantified by the difference between \texttt{Active connections} and \texttt{Strength}). Results displayed in Table \ref{tab:tab2} show an overall prevalence of mobility in non-STEM fields both for male and female. However, the 26\% of male participants study a STEM discipline in 2008 against the 14\% of female participants; the imbalance persists in 2013 with 17\% of males and 11\% of females studying in STEM, revealing also an overall decreasing proportion of Erasmus students with disabilities in STEM disciplines.

\begin{table}[!ht]
\begin{center}
\caption{\small{Summary statistics - Erasmus network Special needs - 2008 and 2013}}
\begin{footnotesize}
\begin{tabular}{lcccccc}
\toprule
\multirow{1}*{} & \multicolumn{3}{c}{\texttt{2008}} & \multicolumn{3}{c}{\texttt{2013}}\\
\cmidrule(l){2-4} \cmidrule(l){5-7}
                        & all    & M      &  F      & all     & M      &  F  \\
\midrule
\multirow{1}*{\texttt{Active Universities}}    & 252 & 130 & 170 & 388 & 198 &  277\\
\multirow{1}*{}                                 & [2290] & & & [2658] & & \\
\multicolumn{1}{r}{\texttt{sending}} & 122 & 65 & 82 & 187 & 97 &  134\\
\multicolumn{1}{r}{\texttt{receiving}}  & 160 & 74 & 104 & 245 & 112 &  167\\[1.0ex]
\multirow{1}*{\texttt{University partnerships}}    & 199  &  &  & 324 &   & \\[0.5ex]
\multirow{1}*{\texttt{Active connections}}    & 202  & 80 & 122 & 328 & 125  & 203 \\[0.5ex]

\midrule
\multirow{1}*{\texttt{Isolates}}   & 649 & 771 & 731 & 513 & 703 & 624\\[0.3ex]

\multirow{1}*{\texttt{Density}}    & 0.0002 & 0.0001 & 0.0002 &  0.0004 & 0.0002 & 0.0003 \\
\multirow{1}*{}    & [0.006] & & & [0.008] & & \\[0.3ex]

\multirow{1}*{\texttt{Degree}}             & 0.004 & 0.002 & 0.003 & 0.005 & 0.003 &  0.003\\
\multicolumn{1}{r}{\texttt{out}}           & 0.008 & 0.003 & 0.007 & 0.008 & 0.004 &  0.006\\
\multicolumn{1}{r}{\texttt{in}}            & 0.004 & 0.003 & 0.003 & 0.005 & 0.002 &  0.004\\[0.3ex]

\multirow{1}*{\texttt{Closeness}}       & 0.00021 & 0.00001 & 0.00006 & 0.00045 & 0.00002 & 0.00012\\
\multicolumn{1}{r}{\texttt{out}}             & 0.00002 & 0 & 0.00001 & 0.00003 & 0.00001 & 0.00002\\
\multicolumn{1}{r}{\texttt{in}}              & 0.000012 & 0 & 0.00001 & 0.00002 & 0 &  0.00001\\[0.3ex]

\multirow{1}*{\texttt{Assortativity}}     & -0.0194 &  -0.016 & -0.0859 & 0.0813 & -0.0636 & 0.1001\\
\midrule
\multirow{1}*{\texttt{Strength}}  & 203 & 80 & 123 & 331 & 125 & 206 \\
\multicolumn{1}{r}{\texttt{STEM}}    & 39 &  21 & 18 & 45 & 22 & 23 \\
\multicolumn{1}{r}{\texttt{non-STEM}}    & 164 & 59 & 105 & 286 & 103 & 183 \\[0.3ex]
\bottomrule

\end{tabular}
\end{footnotesize}
\label{tab:tab2}
\end{center}

{\scriptsize {\bf Note}: {\raggedright \linespread{0.5}\selectfont See \cite{WasFau1994} for the definition of the statistics used. M stands for male; F stands for female. \texttt{Degree} stands for Degree centralization (standardized); \texttt{Closeness} stands for Closeness centralization (standardized); \texttt{Active connections} includes student flows in different fields of study; the \texttt{Assortativity} score is [-1,1]. Values in squared brackets refer to the entire network of Erasmus student flows and are taken from \citep{de2020gender}. \par}} 

\end{table}

\begin{figure}[h!]	
\subfloat[]{%
\includegraphics[width=11cm, height=8.5 cm]{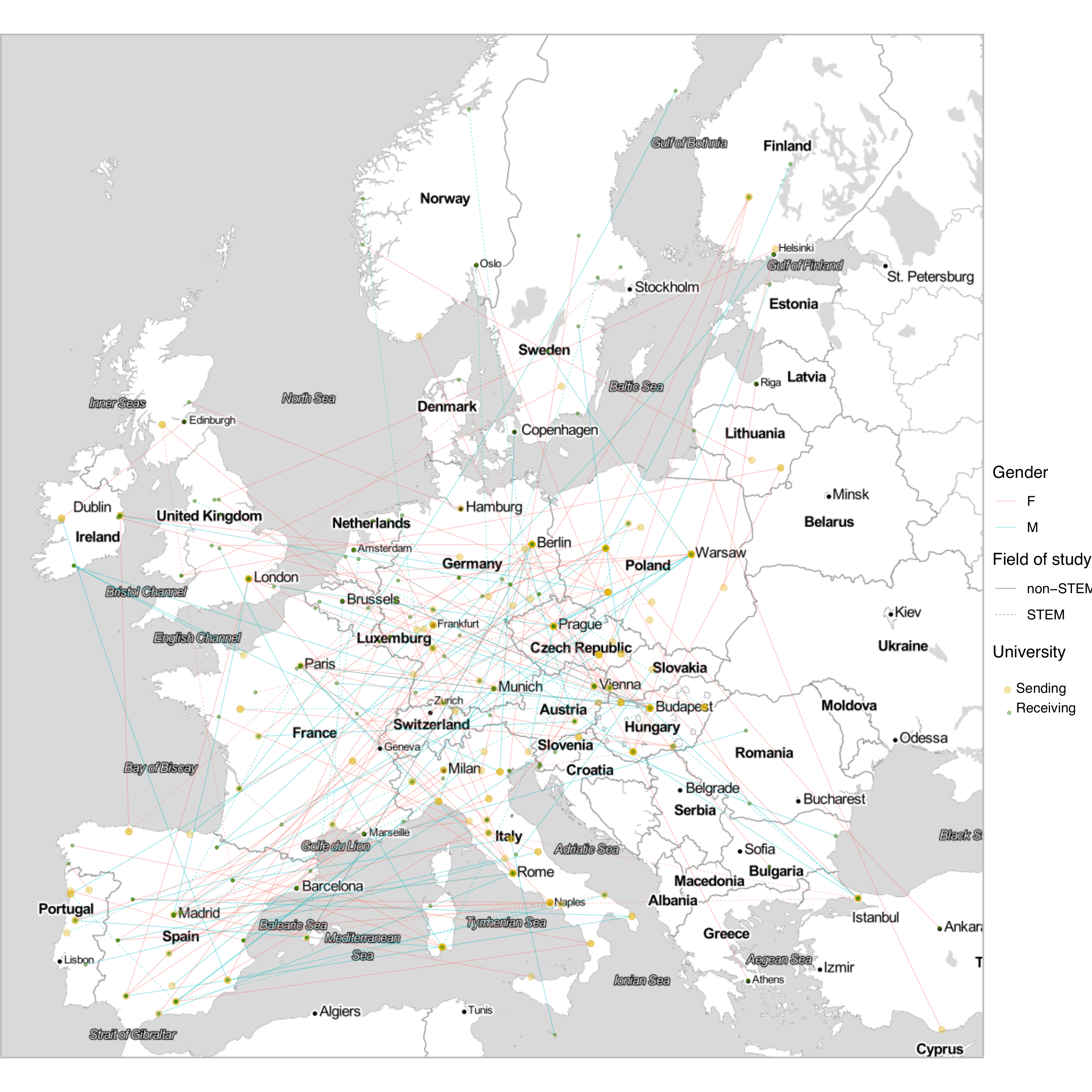}
}

\subfloat[]{%
\includegraphics[width=11cm, height=8.5 cm]{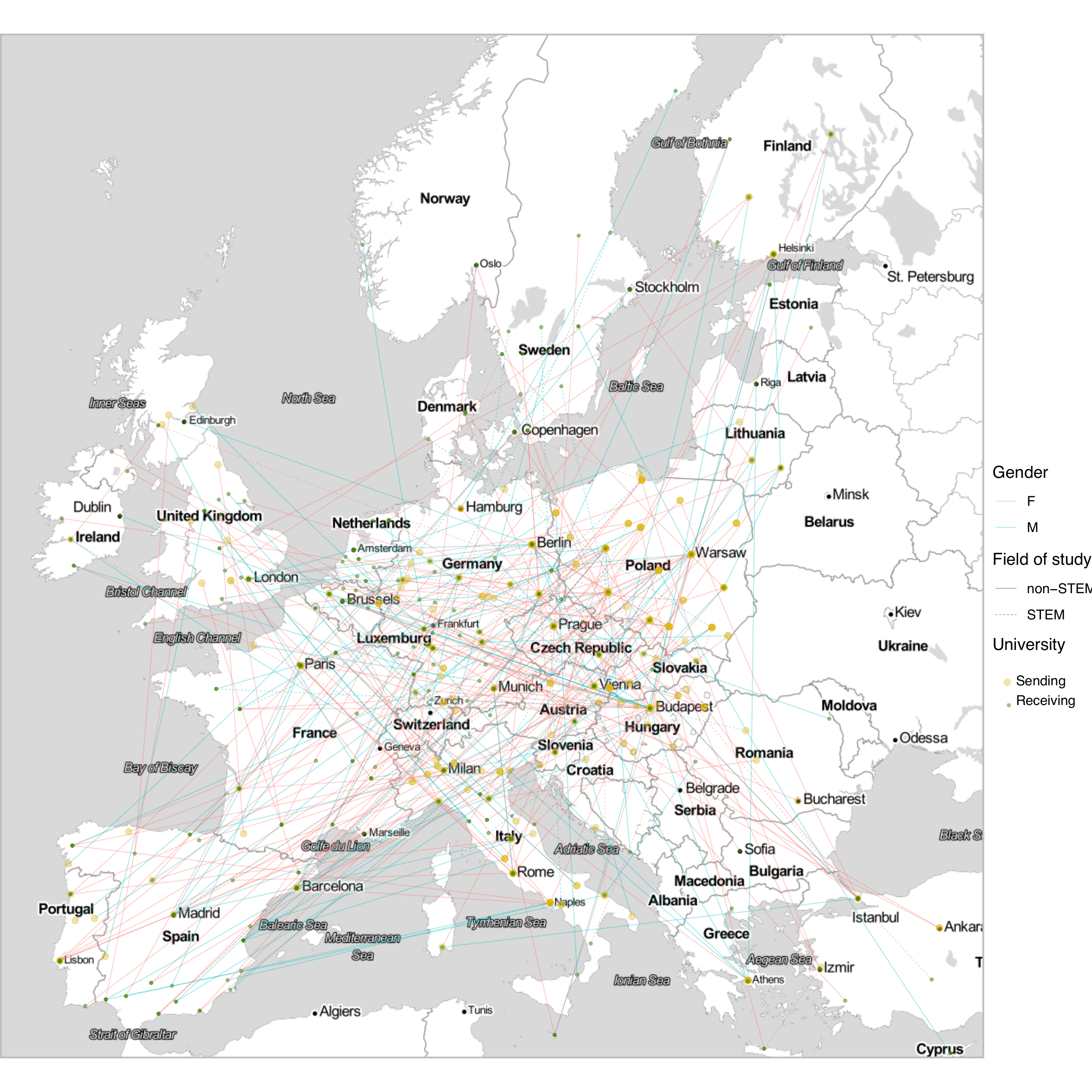}
}
	\caption{The network of Erasmus universities sending and receiving students with special needs in 2008 (a) and 2013 (b).\\
	Yellow (Green) dots represent sending (receiving) universities. Red (Cyan) links represent flows of female (male) Erasmus students. Continuous (Dotted) links represent flows of students enrolled in non-STEM (STEM) fields.}
	\label{fig:fig3}
\end{figure}

\begin{figure}[h!]	
\subfloat[]{%
\includegraphics[width=12.5cm, height=8 cm]{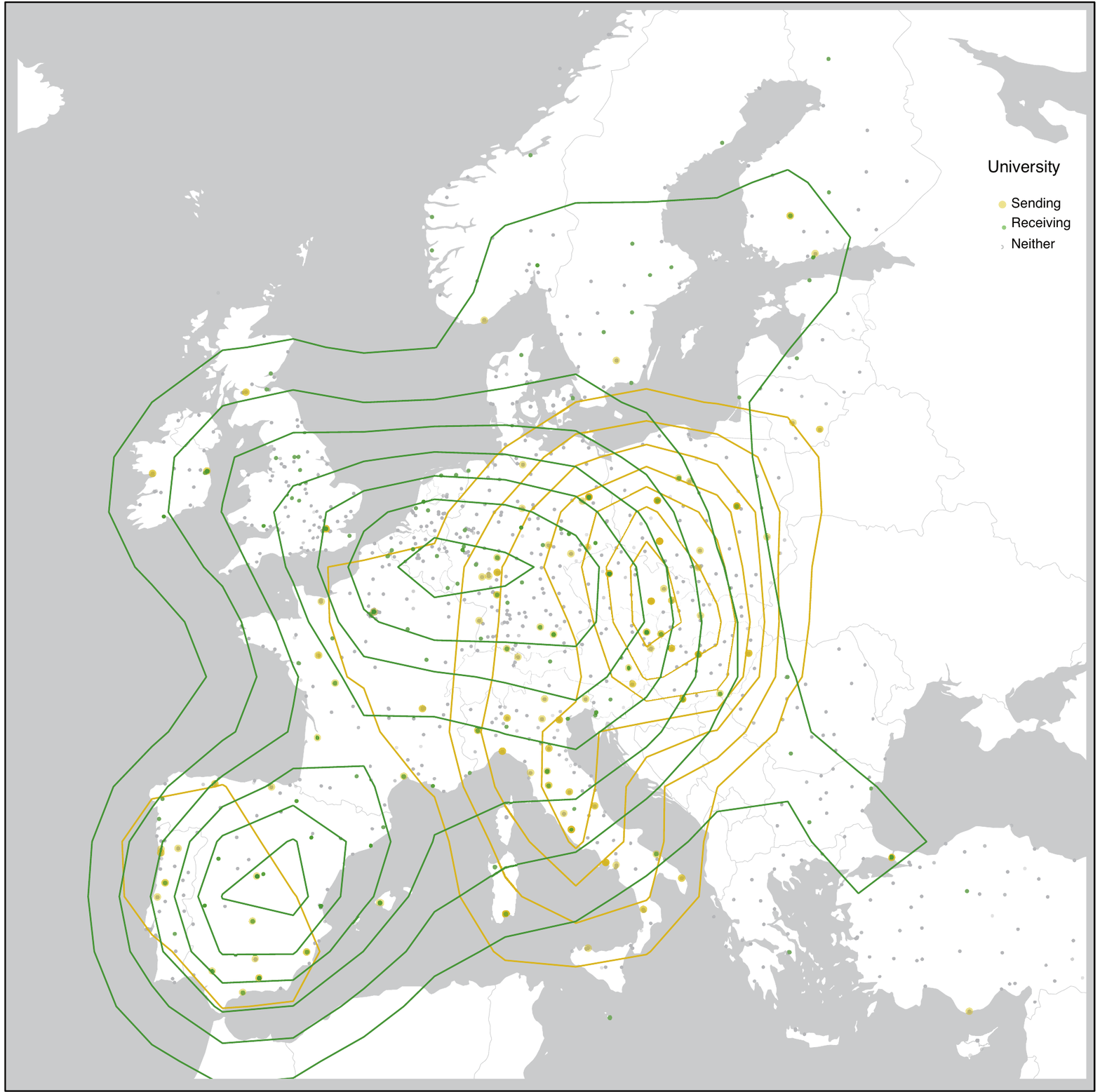}
}

\subfloat[]{%
\includegraphics[width=12.5 cm, height=8 cm]{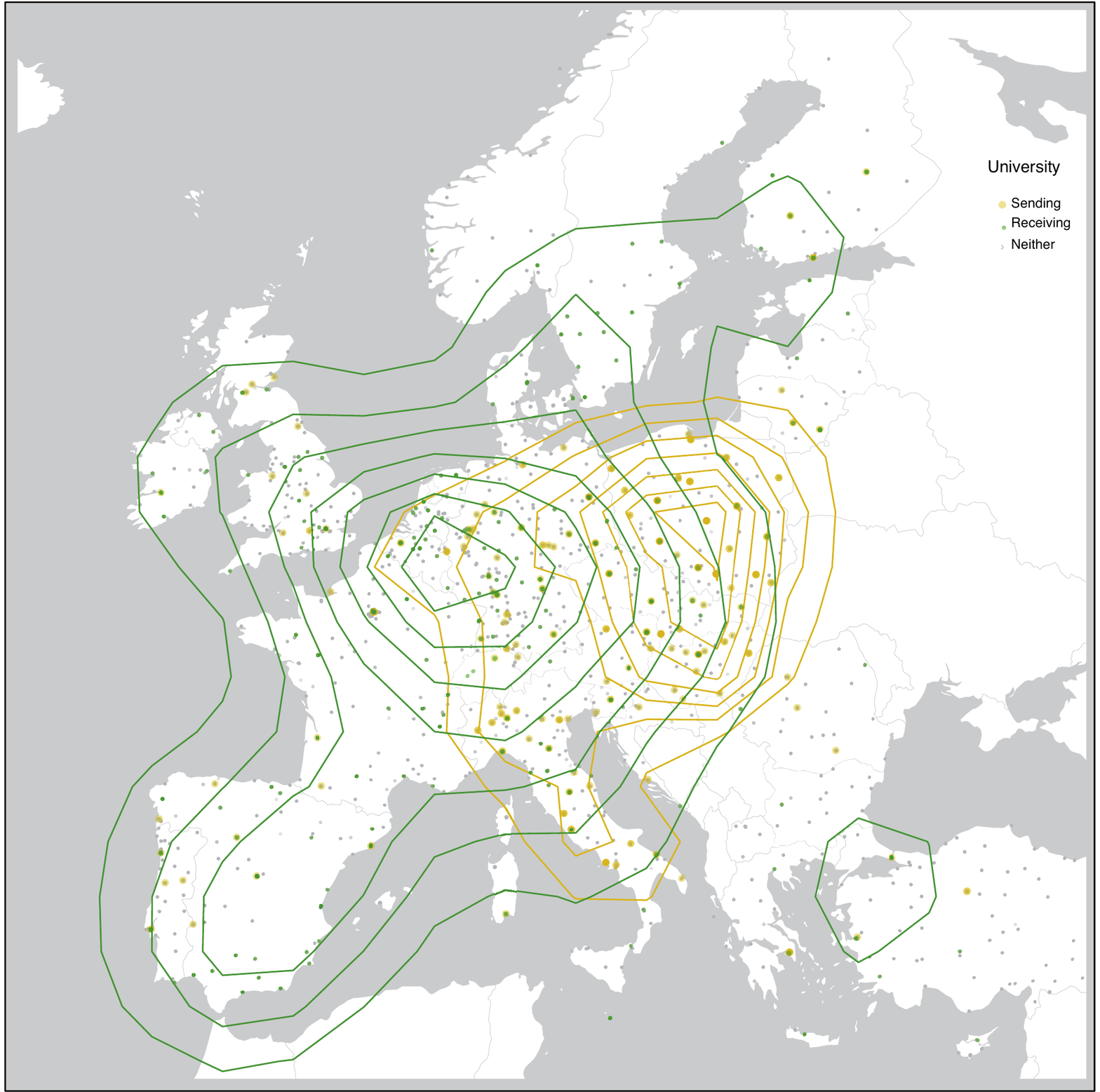}
}
	\caption{The polarization of sending and receiving universities in 2008 (a) and 2013 (b).\\
	Yellow (Green) dots represent sending (receiving) universities. Grey dots represent universities that neither send nor receive Erasmus students with special needs. The Yellow (Green) contour visualizes the density of sending (receiving) universities.}
	\label{fig:fig4}
\end{figure}

The use of Network Analysis allows to give evidence to the structural characteristics of the international flows of Erasmus students with disabilities and, in this case, to acquire information on the connections among universities. It also allows to identify the institutions that play a central role in the mobility program. This information can be acquired through different centrality measures. For instance, the hub and authorities centrality scores \citep{kleinberg1999authoritative} could help identify sending universities (hubs) and receiving universities (authorities), however, the computation of these scores loses relevance in such a sparse network. Thus, we rely on the indegree and outdegree centrality measures to identify the top sending and receiving universities, which are displayed in Table \ref{tab:tab3}. These centrality measures contribute to highlight the difference in connectivity between the female and male graph and its persistence over time. The nodes with the highest outdegree centrality in 2008 are the Adam Mickiewicz University of Poznan in Poland (PL POZNAN01) and the Eötvös Loránd University in Budapest (HU BUDAPES01), Hungary, with $\text{deg}(v_{max})$ equal to 7, and in 2013 two Polish institutions, the Jagiellonian University in Krakow (PL KRAKOW01) and the Adam Mickiewicz University of Poznan, with $\text{deg}(v_{max})$ equal to 8. In the network including non-disabled students \citep{de2020gender}, the centrality measures showed that the top sending universities roughly coincided with the top receiving ones. However, this is not the case in the network of students with disabilities: centrality measures show a polarization of universities, with a different geography for sending and receiving universities. The University of Granada (E GRANADA01) in Spain and the Charles University in Prague (CZ PRAHA07), Czech Republic, are the top receiving universities in 2008 with  $\text{deg}(v_{max})$ equal to 4, whereas the University of Barcelona (E BARCELO01) is the top receiving in 2013 ($\text{deg}(v_{max})= 5$). This polarization becomes more evident in Figures \ref{fig:fig3} and \ref{fig:fig4}. We exploited the information related to the cities where universities are located to extract geographical coordinates through Google Maps and combined the network visualization with the map of Europe. Figure \ref{fig:fig3} shows that the network becomes denser along the years considered, as reported in Table \ref{tab:tab2} and highlights the role of universities as senders and/or receivers respectively in yellow and green. The maps reveal that the universities located in countries capital cities tend to be simultaneously sending and receiving universities, while the rest of institutions mostly exclusively send or host international students, following a precise trajectory, as highlighted by the contour lines in Figure \ref{fig:fig4}. Sending universities (yellow dots) are concentrated (yellow contour) in Italy, Germany and Eastern countries, whereas receiving institutions (green dots) follow a South-West North-East trajectory (green contour), including Spain, the UK and Northern European countries. This polarization and trajectory gets reinforced in 2013. Universities that do not participate to the network of Erasmus students with disabilities are visualized as gray dots and are ten times more numerous.

\begin{table}[!ht]
\begin{center}
\caption{Summary statistics - Erasmus network Special needs - 2008 and 2013}
\begin{footnotesize}
\begin{tabular}{lcccccc}
\toprule
\multirow{1}*{} & \multicolumn{3}{c}{\texttt{2008}} & \multicolumn{3}{c}{\texttt{2013}}\\
\cmidrule(l){2-4} \cmidrule(l){5-7}
                        & all    & M      &  F      & all     & M      &  F  \\
\midrule
\multirow{10}{*}{Top-5 sending}              & POZ01   & WAR01      & POZ01       & KRA01   & BER01        & POZ01\\
            & [7]    &  [3]     & [6]      & [8]          &     [4]    &  [6]\\[0.5ex]
            & BUD01   & PAD01    & BER01       & POZ01         & ATH01        &  WAR01\\
            & [7]     & [2]    & [5]       & [8]  & [3]         &  [5]\\[0.5ex]
        & WAR01     & POR02     & BUD01       & WAR01           & KRA01        &  KRA01\\
universities            & [6]     & [2]       & [5] &  [7]   & [3]        &  [5]\\[0.5ex]
        & BER01     & PER01      & BRN05        & BUD03          & BRA02         &  BUD03\\
    & [6]    & [2]     & [4]        & [7]         & [3]         &  [5]\\[0.5ex]
                                             & WRO01     & CAG01      & WRO01        & BUD20           & RZE02         &  GDA02\\
                                             & [6]     & [2]   & [4]       & [6]          & [3]        &  [4]\\[1ex]
\midrule
\multirow{10}{*}{Top-5 receiving}            & GRA01 & COR01 & MAD14 & BAR01 & MAN04 & BAR01 \\
                & [4] & [3] & [3] & [5] & [2] & [4]\\[0.5ex]
                & PRA07 & GRA01 & MAD03 &  SAN01 & LIN01 & GEN01 \\
                & [4] & [3] & [3] & [4] & [2] & [3]\\[0.5ex]
                 & COR01 & LEI01 & GOT01 & BER01 & AMS02 & SAN01\\
                     universities            & [3] & [2] & [3] & [4] & [2] & [3]\\[0.5ex]
                                             & MAD14 & PRA07 & DUB04 & GEN01 & ANG01 & LEU01\\
                                             & [3] & [2] & [3] & [3] & [2] & [3]\\[0.5ex]
                                             & VAL01 & PRE01 & PON01 & LUN01 & HUE01 & BER01\\
                                             & [3] & [1] & [2] & [3] & [2] & [3]\\
\bottomrule
\end{tabular}
\end{footnotesize}
\label{tab:tab3}
\end{center}

{\scriptsize {\bf Note}: {\raggedright \linespread{0.5}\selectfont Erasmus university codes have been shortened for visualization purpose:          AMS02 = NL AMSTERD02; ANG01 = F ANGERS01; ATH01 = G ATHINE01; BAR01 = E BARCEL01; BER01= D BERLIN01; BRA02 = SK BRATISL02; BRN05 = CZ BRNO05; BUD01= HU BUDAPES01; BUD03 = HU BUDAPES03; BUD20 = HU BUDAPES20; CAG01 = I CAGLIAR01; COR01 = IRLCORK01; DUB04 = IRLDUBLIN04; GDA02 = PL GDANSK02; GEN01 = CH GENEVE01; GOT01 = D GOTTING01;  GRA01 = E GRANADA01; HUE01 = E HUELVA01; KRA01 = PL KRAKOW01; LEI01 = D LEIPZIG01; LEU01 = B LEUVEN01; LIN01 = S LINKOPI01; LUN01 = S LUND01; MAD03 = E MADRID03; MAD14 = E MADRID14; MAN04 = UK MANCHES04; PAD01 = I PADOVA01; PER01 = I PERGUGIA01; PON01= UK PONTYPRO01; POR02 = P PORTO02; POZ01 = PL POZNAN01; PRA07 = CZ PRAHA07; PRE01 = UK PRESTON01; RZE02 = PL RZESZOW02; SAN01 = E SANTIAG01;  VAL01 = E VALENCI01; WAR01 = PL WARSZAW01; WRO01 = PL WROCLAW01.
Squared parentheses contain the degree value.\par}} 

\end{table}

\clearpage
\thispagestyle{empty}

\clearpage
\thispagestyle{empty}

\section{Measuring inclusiveness}

The indegree centrality measures offer a good indication of which universities are the most inclusive in the two reference years, however, they provide an absolute figure, regardless of the general trend of the country they belong to and the total number of incoming Erasmus students. 
In order to obtain a relative measure, we consider the weighted network of universities and propose an index of inclusiveness built as follows:
\[
I^B = \frac{I + 1}{I - 1} \hspace{1cm} \text{where} \hspace{1cm} I = \texttt{instrength}_u \times \frac{i_{c}}{i_{u}} \equiv \frac{i_{sn,u}}{i_{sn,c}} \times \frac{i_{c}}{i_{u}} ,
\]
where $i_{sn,u}$ and $i_{sn,c}$ are respectively the number of incoming students with special needs by university and by country, whereas $i_{u}$ and $i_{c}$ are respectively the number of incoming students, disabled and non-disabled, by university and by country. 
The superscript $B$ stands for \textit{bounded}, as the index is a symmetric transformation of $I$ to obtain a value ranging between [-1,1]. The index represents a measure for the distance between the ratio of incoming students with disabilities at university level and at country level, and quantifies how far the university trends is from the country average. Values equal to zero indicate a perfect alignment of institutions to their country; positive (negative) values indicate a positive (negative) misalignment with respect to their country, i.e. a higher (lower) level of inclusiveness with regard to the average of their country. 

\begin{figure}
\centering
      \includegraphics[scale=0.53]{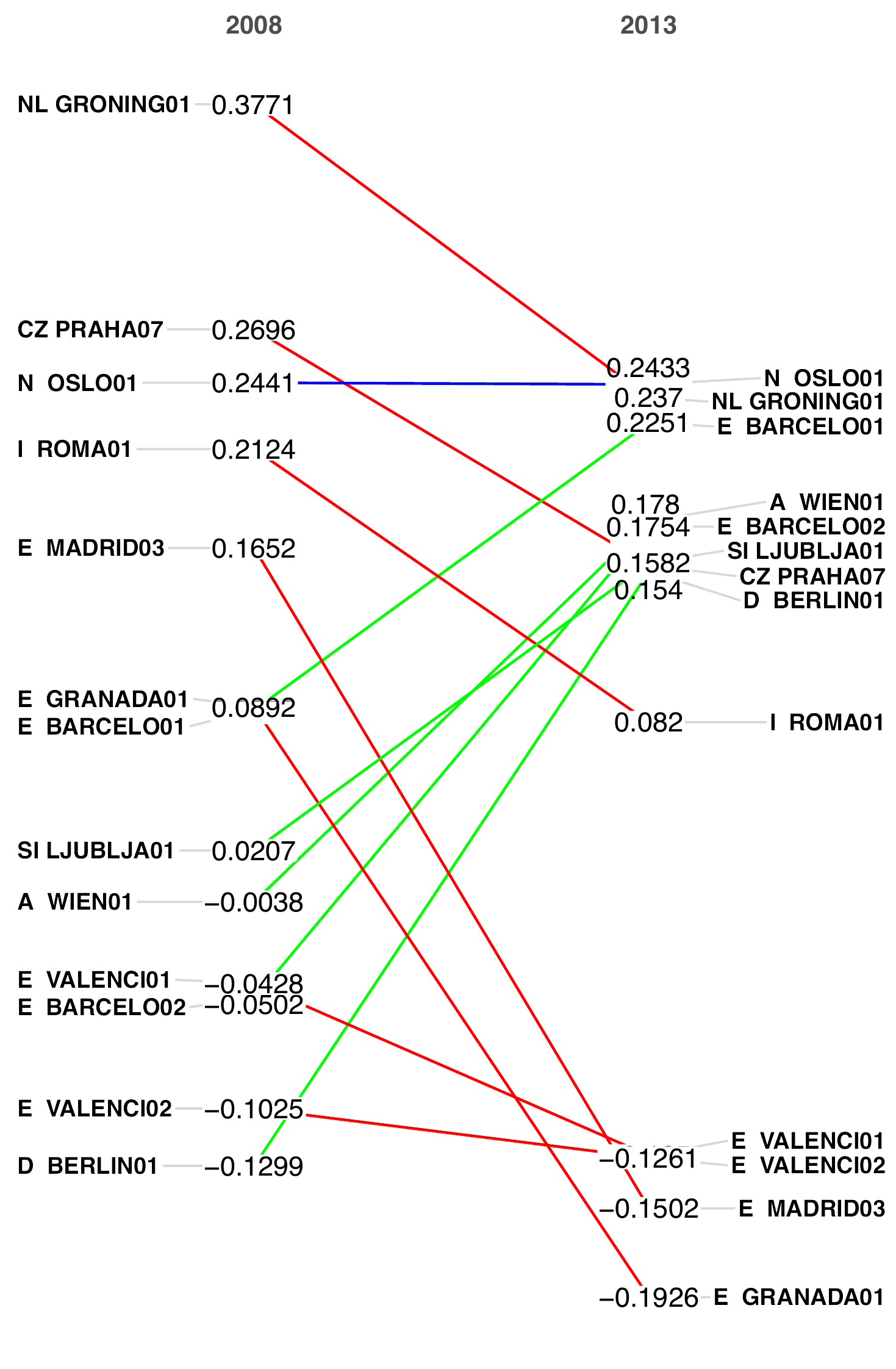}
       \caption{Average index of inclusiveness at the beginning and at the end of the period 2008-2013.}
      \label{fig:fig5}
\end{figure}

To take into account variations along time the index is averaged across 2008, 2009 and 2010 to obtain a mean value for the beginning of the time span, and across 2011, 2012 and 2013 to compute an average value for the end of the period considered. We only consider the universities receiving students with special needs in each of the six years of reference, that is 13 HE institutions. A very small number of universities was able to continuously welcome international students with special needs. The measures obtained are displayed in the Tufte's slopegraph in Figure \ref{fig:fig5}.

Values tend to be close to zero, meaning that the level of inclusiveness of the 13 universities examined does not differ much from the respective country average. At the beginning of the period the Rijksuniversiteit Groningen (NL GRONING01) is on average the most inclusive universities according to our definition, while the Freie Universität Berlin (D BERLIN01) is the least inclusive among the group analyzed. At the end of the period, the University of Oslo and the University of Granada are on average the most and the least inclusive of the group of universities hosting foreign Erasmus students with disabilities.

Three different behaviors can be observed along time. First, the University of Oslo does not undergo any change and its value remains constant. Second, a group of universities worsens its level of inclusiveness, in particular the Universidad Complutense de Madrid (E MADRID03) and the University of Granada (E GRANADA01). These two institutions outperformed the country average level of inclusiveness at the beginning of the period, but fell behind it at the end. Third, a group of universities improves its level of inclusiveness, in particular the Freie Universität Berlin, the Universidad Autonoma de Barcelona (E BARCELO01) and the Universität Wien (A WIEN01) initially show a negative value of the index but eventually outperform their country average.

A further change observable over time is that values are initially more heterogeneous, while at the end of the period they tend to concentrate in two groups: universities outperforming their respective national average level of inclusiveness, and universities which are less inclusive than the country they belong to. The former group include for instance the University of Oslo and the Rijksuniversiteit Groningen, whereas a group of Spanish universities, e.g. the University of Valencia and the Polytechnic University of Valencia, belong to the latter.



\section{Concluding remarks}

The Erasmus Program has been characterized by a low level of participation by students with special needs. Combining different data sources we quantified this evidence and explored the level of inclusiveness and the network of universities contributing to the mobility of students with disabilities. 

The main results of the study can be summarized as follows:
\begin{itemize}
    \item The data collection highlighted a problem of availability and reliability of data on students with special needs in HE in European countries. On the one hand, reasons can be searched in the sensitive nature of data and their self-disclosure requirement; on the other hand, the collection of this information at country level appears still disorganized and in its early stages.
    
    \item Participation in Erasmus by students with disabilities is extremely low with respect to the total number of Erasmus students and to the number of students with disabilities enrolled in HE in Europe. However, the share of Erasmus students with disabilities has doubled over the period 2008-2013.
    
    \item In almost every participating country in 2013, a higher share of female than male students in HE took part in the mobility. The network of universities involved in the mobility of students with special needs show a gender bias in favor of female connections which has been increasing along the period 2008-2013, contrary to the mild tendency towards a reduction of the bias observed in previous research for the network of disabled and non-disabled students.
    
    \item Mobility in the STEM fields has been diminishing over 2008-2013 for both male and female students with disabilities.
    
    \item Universities involved in the mobility of students with special needs tend to polarize in the role of sender or receiver, following a well defined geographical trajectory: universities from the South-West North-East European axis, i.e. those located in Spain, France, UK, the Netherlands and Scandinavian countries, are receiving universities, whereas institutions located in Italy, Germany and Eastern countries are sending universities. The exception is represented by universities located in capital cities, which tend to be both sender and receivers.
    
    \item In the 6-year period between 2008 and 2013 only 13 universities continuously welcomed Erasmus students with disabilities each year. Universities such as the Rijksuniversiteit Groningen and the University of Oslo are among the most inclusive, outperforming their respective national average of incoming students with special needs. The University of Valencia and the Polytechnic University of Valencia are instead among those furthest from reaching the national average. 
    
\end{itemize}

Evidence from this work could be relevant for policy makers defining the Erasmus priorities as well as for European universities as a basis to define an inclusive internal policy and launch appropriate initiatives for better inclusion.
The main limitation of this study consists in the availability and reliability of data, which affects the robustness of our findings. Approximations have been used to study an aspect that would require much greater attention. For this reason, we encourage the European HE institutions to start a systematic collection of information on the participation to education by students with special needs. Quantifying the phenomenon can be a key strategy to widen their participation in HE and international mobility.
Although low, the participation to Erasmus by students with special needs has been growing and it will probably continue to follow this trend in the future, considering the greater budget dedicated to the next seven years of the program. It could be relevant then to further explore its time dimension, analyzing a longer time span and investigating possible similarities or differences across the period through the use of a dynamic network approach \citep{batagelj2014understanding}.
In addition, evidence from this work showed that the UK has a relevant role as receiver of international students with disabilities. Following Brexit, the UK has left the Erasmus program. Future research could explore how this may affect the participation of students with special needs and the probability for them to undertake the mobility in different universities or renounce to the international experience. We assume that this might be also affected by the current pandemic which may constitute a bigger obstacle for students with health issues. Finally, evidence from this work leaves doubts about the reasons why more female than male students with impairments take part in the Erasmus mobility.

\subsection*{List of abbreviations}
ADHD: attention deficit hyperactivity disorder; ERASMUS: EuRopean community Action Scheme for the Mobility of University Students; EU: European Union; HE: Higher Education; ICTs: Information and Communication Technologies; ISCED: International Standard Classification of Education; LDs: learning disabilities; SNA: Social Network Analysis; STEM: Science, Technology, Engineering and Mathematics.

\subsection*{Availability of data and materials}
The original datasets used during the current study are available at the following repositories:
\begin{itemize}
    \item EU open data portal [\url{https://data.europa.eu/en}]
    \item Eurostudent [\url{http://database.eurostudent.eu/}]
    \item Eurostat [\url{https://ec.europa.eu/eurostat/web/education-and-training/data/database}]
    \item Higher Education Statistics Agency [\url{https://www.hesa.ac.uk/data-and-analysis/students/table-15}]
\end{itemize}

The datasets and R scripts generated by the authors are available upon request for replication purposes.

\bibliography{ref}

\begin{thebibliography}{44}
\providecommand{\natexlab}[1]{#1}
\providecommand{\url}[1]{\texttt{#1}}
\expandafter\ifx\csname urlstyle\endcsname\relax
  \providecommand{\doi}[1]{doi: #1}\else
  \providecommand{\doi}{doi: \begingroup \urlstyle{rm}\Url}\fi

\bibitem[Batagelj et~al.(2014)Batagelj, Doreian, Ferligoj, and
  Kejzar]{batagelj2014understanding}
Vladimir Batagelj, Patrick Doreian, Anuska Ferligoj, and Natasa Kejzar.
\newblock \emph{Understanding large temporal networks and spatial networks:
  Exploration, pattern searching, visualization and network evolution},
  volume~2.
\newblock John Wiley \& Sons, 2014.

\bibitem[Behrnd and Porzelt(2012)]{behrnd2012intercultural}
Verena Behrnd and Susanne Porzelt.
\newblock Intercultural competence and training outcomes of students with
  experiences abroad.
\newblock \emph{International Journal of Intercultural Relations}, 36\penalty0
  (2):\penalty0 213--223, 2012.

\bibitem[Bottcher et~al.(2016)Bottcher, Araujo, Nagler, Mendes, Helbing, and
  Herrmann]{bottcher2016gender}
Lucas Bottcher, Nuno~AM Araujo, Jan Nagler, Jose~FF Mendes, Dirk Helbing, and
  Hans~J Herrmann.
\newblock {Gender Gap in the ERASMUS Mobility Program}.
\newblock \emph{PLoS ONE}, 11\penalty0 (2), 2016.

\bibitem[Bound et~al.(2001)Bound, Brown, and Mathiowetz]{bound2001measurement}
John Bound, Charles Brown, and Nancy Mathiowetz.
\newblock Measurement error in survey data.
\newblock In \emph{{Handbook of Econometrics}}, volume~5, pages 3705--3843.
  Elsevier, 2001.

\bibitem[Braskamp et~al.(2009)Braskamp, Braskamp, and
  Merrill]{braskamp2009assessing}
Larry~A Braskamp, David~C Braskamp, and Kelly Merrill.
\newblock Assessing progress in global learning and development of students
  with education abroad experiences.
\newblock \emph{Frontiers: The Interdisciplinary Journal of Study Abroad},
  18:\penalty0 101--118, 2009.

\bibitem[Breznik(2017)]{breznik2017institutional}
Kristijan Breznik.
\newblock {Institutional network of engineering students in the Erasmus
  programme}.
\newblock \emph{Global Journal of Engineering Education}, 19\penalty0
  (1):\penalty0 36--41, 2017.

\bibitem[Breznik and Djakovi{\'c}(2016)]{breznik2016erasmus}
Kristijan Breznik and Goran Djakovi{\'c}.
\newblock Erasmus student mobility flows-the national-level social network
  analysis of slovenia.
\newblock \emph{International Journal of Innovation and Learning}, 20\penalty0
  (2):\penalty0 124--137, 2016.

\bibitem[Breznik and Ragozini(2015)]{breznik2015exploring}
Kristijan Breznik and Giancarlo Ragozini.
\newblock {Exploring the Italian Erasmus agreements by a network analysis
  perspective}.
\newblock In \emph{Proceedings of the 2015 IEEE/ACM International Conference on
  Advances in Social Networks Analysis and Mining 2015}, pages 837--838, 2015.

\bibitem[Breznik and Skrbinjek(2020)]{breznik2020erasmus}
Kristijan Breznik and Vesna Skrbinjek.
\newblock Erasmus student mobility flows.
\newblock \emph{European Journal of Education}, 55\penalty0 (1):\penalty0
  105--117, 2020.

\bibitem[Bry{\l}a(2015)]{bryla2015impact}
Pawe{\l} Bry{\l}a.
\newblock The impact of international student mobility on subsequent employment
  and professional career: A large-scale survey among polish former erasmus
  students.
\newblock \emph{Procedia-Social and behavioral sciences}, 176:\penalty0
  633--641, 2015.

\bibitem[De~Benedictis and Leoni(2020)]{de2020gender}
Luca De~Benedictis and Silvia Leoni.
\newblock {Gender bias in the Erasmus network of universities}.
\newblock \emph{Applied Network Science}, 5\penalty0 (1):\penalty0 1--25, 2020.

\bibitem[de~Boer et~al.(2013)de~Boer, Pijl, Post, and Minnaert]{de2013peer}
Anke de~Boer, Sip~Jan Pijl, Wendy Post, and Alexander Minnaert.
\newblock Peer acceptance and friendships of students with disabilities in
  general education: The role of child, peer, and classroom variables.
\newblock \emph{Social Development}, 22\penalty0 (4):\penalty0 831--844, 2013.

\bibitem[Derzsi et~al.(2011)Derzsi, Derzsy, K{\'a}ptalan, and
  N{\'e}da]{derzsi2011topology}
Aranka Derzsi, Noemi Derzsy, Erna K{\'a}ptalan, and Zolt{\'a}n N{\'e}da.
\newblock Topology of the erasmus student mobility network.
\newblock \emph{Physica A: Statistical Mechanics and its Applications},
  390\penalty0 (13):\penalty0 2601--2610, 2011.

\bibitem[Doyle et~al.(2010)Doyle, Gendall, Meyer, Hoek, Tait, McKenzie, and
  Loorparg]{doyle2010investigation}
Stephanie Doyle, Philip Gendall, Luanna~H Meyer, Janet Hoek, Carolyn Tait,
  Lynanne McKenzie, and Avatar Loorparg.
\newblock An investigation of factors associated with student participation in
  study abroad.
\newblock \emph{Journal of studies in international education}, 14\penalty0
  (5):\penalty0 471--490, 2010.

\bibitem[du~Toit(2018)]{du2018designing}
Nina~HG du~Toit.
\newblock {Designing a model for facilitating the inclusion of higher education
  international students with disabilities in South Africa}.
\newblock \emph{Social Inclusion}, 6\penalty0 (4):\penalty0 168--181, 2018.

\bibitem[d’Hombres and Schnepf(2021)]{d2021international}
B{\'e}atrice d’Hombres and Sylke~V Schnepf.
\newblock International mobility of students in italy and the uk: does it pay
  off and for whom?
\newblock \emph{Higher Education}, pages 1--22, 2021.

\bibitem[Engel(2010)]{engel2010impact}
Constanze Engel.
\newblock The impact of erasmus mobility on the professional career: Empirical
  results of international studies on temporary student and teaching staff
  mobility.
\newblock \emph{Belgeo. Revue belge de g{\'e}ographie}, \penalty0 (4):\penalty0
  351--363, 2010.

\bibitem[{European Commission}(2015)]{european2015erasmus+}
{European Commission}.
\newblock Erasmus+ programme: Annual report 2014, 2015.

\bibitem[{European Commission}(2021)]{EuroCommission}
{European Commission}, 2021.
\newblock
  \url{https://ec.europa.eu/commission/presscorner/detail/en/ip_21_1326}.
  Accessed 22 April 2021.

\bibitem[{European Parliament and Council of European
  Union}(2013)]{eu-1288-2013}
{European Parliament and Council of European Union}.
\newblock Regulation (eu) no 1288/2013, 2013.
\newblock
  \newline\url{https://eur-lex.europa.eu/legal-content/EN/TXT/PDF/?uri=CELEX:32013R1288&from=EN}.

\bibitem[{Eurostat}(2016)]{disabilitystat}
{Eurostat}.
\newblock Disability statistics - prevalence and demographics, 2016.
\newblock
  \newline\url{https://ec.europa.eu/eurostat/statistics-explained/pdfscache/34409.pdf}.

\bibitem[{Eurostat}(2018)]{disabilitystatedu}
{Eurostat}.
\newblock Disability statistics - access to education and training, 2018.
\newblock
  \newline\url{https://ec.europa.eu/eurostat/statistics-explained/pdfscache/34423.pdf}.

\bibitem[Farmer et~al.(1999)Farmer, Van~Acker, Pearl, and
  Rodkin]{farmer1999social}
Thomas~W Farmer, Richard~M Van~Acker, Ruth Pearl, and Philip~C Rodkin.
\newblock Social networks and peer-assessed problem behavior in elementary
  classrooms: Students with and without disabilities.
\newblock \emph{Remedial and Special Education}, 20\penalty0 (4):\penalty0
  244--256, 1999.

\bibitem[Hameister et~al.(1999)Hameister, Mathews, Hosley, and
  Groff]{hameister1999college}
Brenda Hameister, Peter Mathews, Nathaniel Hosley, and Margo~Coffin Groff.
\newblock College students with disabilities and study abroad: Implications for
  international education staff.
\newblock \emph{Frontiers: The interdisciplinary journal of study abroad},
  5\penalty0 (2):\penalty0 81--100, 1999.

\bibitem[Heirweg et~al.(2020)Heirweg, Carette, Ascari, and
  Van~Hove]{heirweg2020study}
Sofie Heirweg, Lieve Carette, Andrea Ascari, and Geert Van~Hove.
\newblock Study abroad programmes for all? barriers to participation in
  international mobility programmes perceived by students with disabilities.
\newblock \emph{International Journal of Disability, Development and
  Education}, 67\penalty0 (1):\penalty0 73--91, 2020.

\bibitem[Johnstone and Edwards(2020)]{johnstone2020accommodations}
Christopher Johnstone and Paul Edwards.
\newblock Accommodations, accessibility, and culture: Increasing access to
  study abroad for students with disabilities.
\newblock \emph{Journal of Studies in International Education}, 24\penalty0
  (4):\penalty0 424--439, 2020.

\bibitem[Keogh and Russel-Roberts(2009)]{keogh2009exchange}
Johannes Keogh and Eileen Russel-Roberts.
\newblock Exchange programmes and student mobility: Meeting student’s
  expectations or an expensive holiday?
\newblock \emph{Nurse Education Today}, 29\penalty0 (1):\penalty0 108--116,
  2009.

\bibitem[Kitsantas(2004)]{kitsantas2004studying}
Anastasia Kitsantas.
\newblock Studying abroad: The role of college students' goals on the
  development of cross-cultural skills and global understanding.
\newblock \emph{College Student Journal}, 38\penalty0 (3):\penalty0 441, 2004.

\bibitem[Kleinberg(1999)]{kleinberg1999authoritative}
Jon~M Kleinberg.
\newblock Authoritative sources in a hyperlinked environment.
\newblock In \emph{Journal of the ACM}, volume~46, page 604–632, 1999.

\bibitem[Langley and Breese(2005)]{langley2005interacting}
Crolyn~S Langley and Jeffrey~R Breese.
\newblock Interacting sojourners: A study of students studying abroad.
\newblock \emph{The Social Science Journal}, 42\penalty0 (2):\penalty0
  313--321, 2005.

\bibitem[Maiworm(2001)]{maiworm2001erasmus}
Friedhelm Maiworm.
\newblock Erasmus: continuity and change in the 1990s.
\newblock \emph{European Journal of Education}, 36\penalty0 (4):\penalty0
  459--472, 2001.

\bibitem[Mamas et~al.(2020{\natexlab{a}})Mamas, Bjorklund~Jr, Daly, and
  Moukarzel]{mamas2020friendship}
Christoforos Mamas, Peter Bjorklund~Jr, Alan~J Daly, and Sara Moukarzel.
\newblock Friendship and support networks among students with disabilities in
  middle school.
\newblock \emph{International Journal of Educational Research}, 103:\penalty0
  101608, 2020{\natexlab{a}}.

\bibitem[Mamas et~al.(2020{\natexlab{b}})Mamas, Schaelli, Daly, Navarro, and
  Trisokka]{mamas2020employing}
Christoforos Mamas, Giovanna~Hartmann Schaelli, Alan~J Daly, Henar~R Navarro,
  and Lambri Trisokka.
\newblock Employing social network analysis to examine the social participation
  of students identified as having special educational needs and disabilities.
\newblock \emph{International Journal of Disability, Development and
  Education}, 67\penalty0 (4):\penalty0 393--408, 2020{\natexlab{b}}.

\bibitem[Oborune(2013)]{oborune2013becoming}
Karina Oborune.
\newblock Becoming more european after erasmus? the impact of the erasmus
  programme on political and cultural identity.
\newblock \emph{Epiphany}, 6\penalty0 (1), 2013.

\bibitem[Otero and McCoshan(2006)]{otero2006survey}
Manuel~Souto Otero and Andrew McCoshan.
\newblock Survey of the socio-economic background of erasmus students - dg eac
  01/05. final report.
\newblock 2006.

\bibitem[Papatsiba(2005)]{papatsiba2005political}
Vassiliki Papatsiba.
\newblock {Political and Individual Rationales of Student Mobility: a
  case-study of ERASMUS and a French regional scheme for studies abroad}.
\newblock \emph{European journal of education}, 40\penalty0 (2):\penalty0
  173--188, 2005.

\bibitem[Parey and Waldinger(2010)]{parey2010studying}
Matthias Parey and Fabian Waldinger.
\newblock Studying abroad and the effect on international labour market
  mobility: evidence from the introduction of erasmus.
\newblock \emph{The Economic Journal}, 121\penalty0 (551):\penalty0 194--222,
  2010.

\bibitem[Restaino et~al.(2020)Restaino, Vitale, and
  Primerano]{restaino2020analysing}
Marialuisa Restaino, Maria~Prosperina Vitale, and Ilaria Primerano.
\newblock {Analysing International Student Mobility Flows in Higher Education:
  A Comparative Study on European Countries}.
\newblock \emph{Social Indicators Research}, pages 1--19, 2020.

\bibitem[{Sapienza University of Rome}(2021)]{Sapienza}
{Sapienza University of Rome}, 2021.
\newblock
  \url{https://www.uniroma1.it/en/notizia/sapienza-awarded-honorary-doctorate-sofia-corradi}.
  Accessed 23 April 2021.

\bibitem[Shames and Alden(2005)]{shames2005impact}
Wendy Shames and Peg Alden.
\newblock The impact of short term study abroad on the identity development of
  college students with learning disabilities and/or ad/hd.
\newblock \emph{Frontiers: The Interdisciplinary Journal of Study Abroad},
  11\penalty0 (1):\penalty0 1--31, 2005.

\bibitem[Shields(2013)]{shields2013globalization}
Robin Shields.
\newblock Globalization and international student mobility: A network analysis.
\newblock \emph{Comparative Education Review}, 57\penalty0 (4):\penalty0
  609--636, 2013.

\bibitem[Souto-Otero et~al.(2013)Souto-Otero, Huisman, Beerkens, De~Wit, and
  Vuji{\'c}]{souto2013barriers}
Manuel Souto-Otero, Jeroen Huisman, Maarja Beerkens, Hans De~Wit, and
  Sun{\v{c}}ica Vuji{\'c}.
\newblock {Barriers to international student mobility: Evidence from the
  Erasmus program}.
\newblock \emph{Educational researcher}, 42\penalty0 (2):\penalty0 70--77,
  2013.

\bibitem[Teichler and Jahr(2001)]{teichler2001mobility}
Ulrich Teichler and Volker Jahr.
\newblock Mobility during the course of study and after graduation.
\newblock \emph{European journal of education}, 36\penalty0 (4):\penalty0
  443--458, 2001.

\bibitem[Wasserman and Faust(1994)]{WasFau1994}
Stanley Wasserman and Katherine Faust.
\newblock \emph{{Social Network Analysis: Methods and Applications}}.
\newblock Cambridge University Press, 1994.

\end{thebibliography}

\end{document}